\documentclass[preprint]{article}
\usepackage{epsfig}
\usepackage{authblk}
\usepackage[english]{babel}
\usepackage{graphicx}
\usepackage{bm}
\usepackage{textcmds}
\usepackage{amsmath}
\usepackage{amsfonts}
\usepackage{amssymb}
\usepackage{natbib}
\bibpunct{(}{)}{,}{a}{}{;}

\def\figfigincl#1#2#3{\includegraphics[width=#1]{figures/#2.eps}%
    \caption{\small #3}\label{fig:#2}}
\def\figfiginclnocap#1#2{\includegraphics[width=#1]{figures/#2.eps}}
\newcommand{\R}{\mathbb{R}}
\begin{document}

\title{Nongravitational perturbations and virtual impactors: the case
  of asteroid (410777) 2009~FD}
\author[1]{Federica Spoto\thanks{spoto@mail.dm.unipi.it}}
\author[1]{Andrea Milani}
\author[2]{Davide Farnocchia}
\author[2]{Steven R. Chesley}
\author[3,5]{Marco Micheli}
\author[4,5]{Giovanni B. Valsecchi}
\author[6]{Davide Perna}
\author[7]{Olivier Hainaut}

\affil[1]{Department of Mathematics, University of Pisa,
  Pisa, Italy}
\affil[2]{Jet Propulsion Laboratory, California Institute of Technology, 
  Pasadena, California, USA}
\affil[3]{ESA NEO Coordination Centre, Frascati, Roma,
  Italy}
\affil[4]{IAPS-INAF, Roma, Italy}
\affil[5]{IFAC-CNR, Sesto Fiorentino, Firenze, Italy}
\affil[6]{LESIA - Observatory of Paris, CNRS, UPMC, University of
  Paris-Diderot, Meudon, France}
\affil[7]{European Southern Observatory, Munich, Germany}

\date{Accepted by Astronomy and Astrophysics on September 17, 2014}

\maketitle

\abstract{Asteroid (410777) 2009~FD could hit Earth between 2185 and
  2196. The long term propagation to the possible impacts and the
  intervening planetary encounters make 2009~FD one of the most
  challenging asteroids in terms of hazard assessment. To compute
  accurate impact probabilities we model the Yarkovsky effect by using
  the available physical characterization of 2009~FD and general
  properties of the near Earth asteroid population. We perform the
  hazard assessment with two independent methods: the first method is
  a generalization of the standard impact monitoring algorithms in use
  by NEODyS and Sentry, while the second one is based on a Monte Carlo
  approach.  Both methods generate orbital samples in a
  seven-dimensional space that includes orbital elements and the
  parameter characterizing the Yarkovsky effect. The highest impact
  probability is $2.7 \times 10^{-3}$ for an impact during the 2185
  Earth encounter. Impacts after 2185 corresponding to resonant
  returns are possible, the most relevant being in 2190 with a
  probability of $3 \times 10^{-4}$. Both numerical methods can be
  used in the future to handle similar cases. The structure of
  resonant returns and the list of the possible keyholes on the target
  plane of the scattering encounter in 2185 can be predicted by an
  analytic theory.}



\section{Introduction}
\label{sec:intro}

For many years we have been operating impact monitoring
  systems at the University of
Pisa\footnote{http;//newton.dm.unipi.it/neodys since 1999; operated by
  SpaceDyS srl. from 2011.} and the Jet Propulsion Laboratory
(JPL)\footnote{http://neo.jpl.nasa.gov/risk/ since 2002.}. These
online information systems continually and automatically update the
list of asteroids that can hit our planet in the next 100 years.

The attempt to extend the monitoring time span to a longer interval,
e.g., 200 years, is on the contrary at the frontier of research on the
theory of chaos, nongravitational perturbations, and observational
error models. Thus, we are not surprised to find that new cases need
to be handled in a different way from the previous ones. So far we
have successfully handled the special cases (99942) Apophis
\citep{apophis}, (101955) Bennu \citep{bennu}, and 1950 DA
\citep{1950DA}. Each of these cases required us to model and/or solve
for parameters appearing in the nongravitational perturbations,
especially the Yarkovsky effect \citep{yarkopred}.

Recently, asteroid 2009~FD (discovered by the La Sagra survey on 2009
March 16) appeared as a new case with the following new
characteristics. We previously had 182 optical observations (from the
years 2009 and 2010) and a very precise orbit solution, with a purely
gravitational model, leading to several Virtual Impactors (VIs)
(patches of initial conditions leading to possible impacts with Earth
\citep{clomon2}) in the years 2185--2196. 2009~FD was reobserved
between 2013 November and 2014 April: 109 additional optical
observations were obtained, plus one radar Doppler measurement was
performed on April 7 from Arecibo (see
Sec.~\ref{sec:observations}). As a consequence, the uncertainty of the
orbit with the same model become small enough to exclude the main VI
in 2185, the one with largest Impact Probability (IP).

However, this result was inaccurate because it did not properly
account for the uncertainties of the dynamical model. The available
astrometry, even with the radar data point, is not sufficient to
determine the strength of the Yarkovsky effect. The Yarkovsky effect
order of magnitude, as estimated by models, increases the uncertainty
of the long term prediction and therefore the main VI in 2185 is still
within the range of possible orbits.

If new observations are added without modeling the Yarkovsky effect,
it is possible that no VIs will be included, although we know this is
not correct. Therefore, rather than removing the risk file
(list of VIs) we need to be able to compute a risk file taking fully
into account the Yarkovsky effect. Otherwise the observers would
decrease the priority of observing 2009~FD. To solve this problem we
started an intensive effort to compute the appropriate solution; in
the meantime we decided not to update the online risk
files\footnote{This decision was applied both at University of
  Pisa/SpaceDyS and at JPL.} to avoid giving a false ``all clear''.

In this paper we report how we solved this problem, in two
different ways, in Pisa and at JPL. Both solutions use theories, most
of which are presented in the papers cited, but some are new, and the
known tools have to be combined in an innovative way to solve this
specific case. Of course our hope is to have accumulated enough
expertise (and well-tested software) to be able to handle new
difficult cases, but this is yet to be determined.

The computation of a Yarkovsky model is based on the available
physical properties of 2009~FD, as well as general properties of the
Near Earth Asteroid (NEA) population, with uncertainties propagated
nonlinearly to generate a Probability Density Function (PDF) for the
Yarkovsky parameter $A_2$ \citep{yarko_list} (see
Sec.~\ref{sec:yarko}). We used this model in two different ways.

The Pisa solution is to generalize the method of the Line Of
Variations (LOV) \citep[]{multsol,clomon2} already in use (both in Pisa
and at JPL) to a higher dimensional space, e.g., to vectors containing
six initial conditions and at least one nongravitational parameter. We
obtained the appropriate metric for defining the LOV by mapping on the
2185 scatter plane (Sec.~\ref{sec:lov}). We control the weakness in
the determination of the Yarkovsky parameter by adding an a priori
observation (Sec.~\ref{sec:apriori}). The JPL solution is based on a
Monte Carlo method applied to propagate the orbital PDF (including the
Yarkovsky parameter) to the target planes of the encounters
with Earth in the late 22nd Century (Sec.~\ref{sec:montecarlo}).

In Sec.~\ref{sec:scattering} we discuss the role of the 2185 close
approach in scattering the alternative orbits and consequently in
giving access to resonant returns. The analytic theory, based on
\cite{analytic}, provides approximate locations for the possible
keyholes is given in Appendix~\ref{sec:appendix}.

The results obtained by the two methods are compared in
Sec.~\ref{sec:compare}, where we dicuss the trade-off between the
two. We also discuss the future observability of 2009~FD.

\section{Astrometry and physical observations of 2009~FD}
\label{sec:observations}
The observational coverage of 2009~FD available to date is composed of
three separate apparitions.  More than 150 astrometric positions were
reported during its discovery apparition in 2009, when 2009~FD reached
a magnitude of V=16 just before disappearing into solar conjunction,
making it an easy target for many observers.  A slightly less
favorable opportunity in late 2010 resulted in a handful of additional
observations, including a Near Infrared (NIR) detection by
the WISE spacecraft \citep{wise}. The object then entered a phase of
almost prohibitive observational geometry, which
resulted in a lack of coverage for a three year period, until late
2013.

In an effort to secure the maximum observational coverage for this
important target, in November 2013 we decided to attempt an early
third-opposition recovery using the 8.2 meter ESO Very Large Telescope
(VLT) on Cerro Paranal, Chile. Observations collected starting from
2013 November 30 with the FORS2 optical imager resulted in a faint but
unambiguous detection inside the uncertainty region, confirmed by
consistent detections achieved over the two subsequent nights; at that
time the object was estimated to have a magnitude of approximately
V=25.5, making it a challenging target even for a large aperture
telescope like VLT. From early 2014 various other professional and
amateur-level sites began reporting optical observations, guaranteeing
a dense astrometric coverage until early April, when the object
reached its close approach with Earth and then entered solar
conjunction.

As an additional attempt to extend the observational coverage, we
tried to locate unreported precovery observations of 2009~FD in
existing archival data, using the image search engine made available
by the Canadian Astronomy Data Centre \citep{gwyn}; all the available
images covering the ephemeris position of 2009~FD corresponded to
times when the object was fainter than V=24, unlikely to result in a
detection in non-targeted sidereal exposures.

\begin{figure}[h]
  \begin{center}
    \includegraphics[width=10cm]{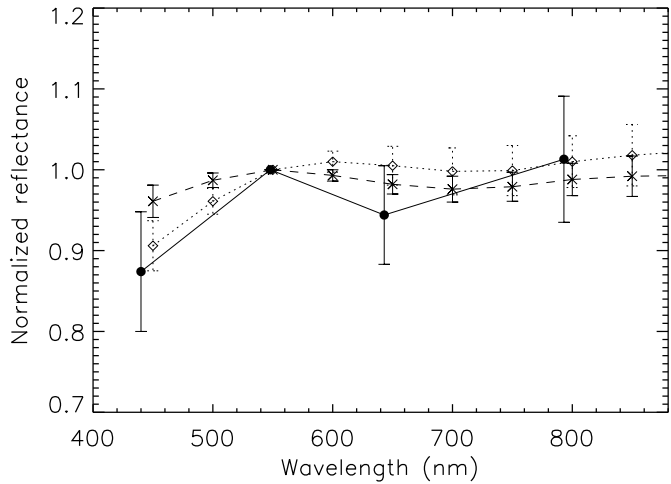}
    \caption{Comparisons of the colors of 2009~FD
      with the visible spectral shapes of the Ch and Cgh
      classes. Continous line: 2009~FD measurements; dashed: Ch
      taxonomic class; dotted: Cgh class.}
    \label{fig:2009FD_tax}
  \end{center}
\end{figure}

Just before the end of the observability window, and close to the time
of peak brightness for the apparition, we were able to obtain BVRI
colorimetric observations using the EFOSC2 instrument mounted on the
3.6 meter ESO New Technology Telescope (NTT) at La Silla, Chile.  The
exposure time was of 200 s for each of the images, which were reduced
using standard procedures with the MIDAS software: after subtraction
of the bias from the raw data and flat-field correction, the
instrumental magnitudes were measured via aperture photometry. For the
R filter, we considered the mean value of two different images, while
only one image was taken with the other photometric filters.  The
absolute calibration of the magnitudes was obtained by means of the
observation of standard fields from the \citep{landolt} catalog.
Although exposed at high airmass (around 1.9) and under not ideal (but
stable) seeing conditions (1.4"), the dataset was sufficient to
extract accurate optical colors for the asteroid (see
Table~\ref{tab:mag_col}), which suggest a C-group primitive
composition, most likely (based on chi-square minimization) of the Ch
or Cgh classes \citep{demeo} (see Fig.~\ref{fig:2009FD_tax}). These
observations were obtained only a few days before the radar Doppler
detection by the Arecibo radiotelescope, which marked the end of the
2013-2014 apparition of 2009~FD.

\begin{table}[ht]
  \footnotesize
  \centering
  \caption{Apparent V magnitude and optical colors (with error bars) of
    2009~FD on 2014 April 02.0 UT. They are consistent with a primitive
    C-group taxonomy, most likely of the Ch or Cgh classes.}
  \label{tab:mag_col}
  \medskip
  \begin{tabular}{lr}
    \hline
    Band & Value [mag]\\
    \hline
    V      & $20.258 \pm 0.063$\\
    B -- V & $ 0.816 \pm 0.091$\\
    V -- R & $ 0.298 \pm 0.070$\\
    V -- I & $ 0.704 \pm 0.083$\\
    \hline
  \end{tabular}
\end{table}

\section{Yarkovsky effect models}
\label{sec:yarko}
As already discussed, the Yarkovsky effect \citep{yarkopred} needs to
be taken into account to make reliable impact predictions for
2009~FD. Including the Yarkovsky accelerations in the force model is
tricky because such accelerations are unknown.

One way to constrain the Yarkovsky effect is to look for deviations
from a gravitational trajectory in the astrometric dataset. The
Yarkovsky effect is modeled as a purely transverse acceleration
$A_2/r^2$ and $A_2$ is determined by the orbital fit to the
observations \citep{yarko_list}. \citep{bennu} successfully used this
approach for asteroid (101955) Bennu. However, for 2009~FD we have a
relatively short observed arc and only one Doppler radar
observation. Therefore, the astrometry provides no useful constraint
on $A_2$.

Another option is to use the available physical model as well as
general properties of the near-Earth asteroid population to constrain
the Yarkovsky effect. \citep{apophis} and \citep{1950DA} applied this
technique to perform the risk assessment of asteroids (99942) Apophis
and (29075) 1950~DA. The situation for 2009~FD is similar to
that discussed by \citep{apophis} for Apophis. The available
information for 2009~FD is as follows.

\begin{itemize}
\item \citep{wise} use WISE observations to constrain the diameter and
  albedo of 2009~FD as $(472 \pm 45)$ m and $(0.010 \pm 0.003)$,
  respectively. This value of the albedo is extreme, lower by a factor
  of $>3$ than any other known albedo for asteroids of similar
  taxonomic classes. Such a large anomaly cannot be due to the error
  in absolute magnitude, thus even the diameter could be
  unreliable. We use the published data: when better data is
  available we can easily repeat the procedure described in this
  paper.
\item The known rotation period is $(5.9 \pm 0.2)$ h \citep{rotP}.
\end{itemize}

\begin{figure}[h]
  \begin{center}
    \includegraphics[width=10 cm]{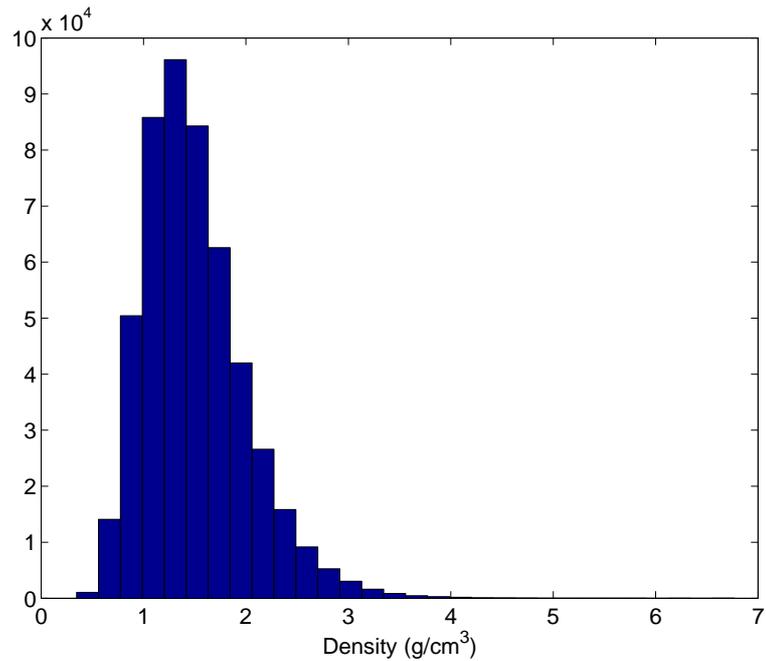}
    \caption{Assumed distribution of the 2009~FD
      density.}
    \label{fig:density_hist}
  \end{center}
\end{figure}

The slope parameter $G$, the obliquity, density, and thermal inertia
are unknown. Therefore, we resort to general properties of the
asteroid population:
\begin{itemize}
\item From the JPL small-body
  database\footnote{http://ssd.jpl.nasa.gov/sbdb\_query.cgi}, we
  obtain $G = (0.18 \pm 0.13)$ for the whole asteroid population. This
  distribution for $G$ was also used by \citep{2009bd} for asteroid
  2009~BD.
\item For the spin axis orientation we use the obliquity distribution
  by \citep{yarko_list}, which was obtained from a list of Yarkovsky
  detections.
\item The density is unknown, but as discussed in
  Sec.~\ref{sec:observations} spectral properties suggest a C-type
  asteroid and therefore a density typically smaller than 2 g/cm$^3$. We used a
  distribution as in Fig.~\ref{fig:density_hist}, i.e., a lognormal
  with mean 1.5 g/cm$^3$ and standard deviation 0.5 g/cm$^3$.
\item For thermal inertia we adopt the \citep{thermal_inertia}
  relationship between diameter and thermal inertia.
\end{itemize}
For more details see \citep{yarko_list} and \citep{apophis}.

Figure~\ref{fig:A2_dist} shows the distribution of $A_2$ obtained by
combining the physical parameters described above. Since we do not
know whether 2009~FD is a retrograde or a direct rotator, the $A_2$
distribution has a bimodal behavior. In general, a retrograde rotation
is more likely as discussed in \citep{laspina} and
\citep{yarko_list}. We did not model a complex rotation
state. However, the overall uncertainty is well captured since
a complex rotation would decrease the size of the Yarkovsky effect and
thus $A_2$, thereby providing no wider
dispersion. Figure~\ref{fig:A2_dist} also shows a normal distribution
with zero mean and the same $3\sigma$ level of the distribution
obtained from the physical model, i.e., $97.5 \times 10^{-15}$
au/d$^2$.

\begin{figure}[t]
\figfigincl{10 cm}{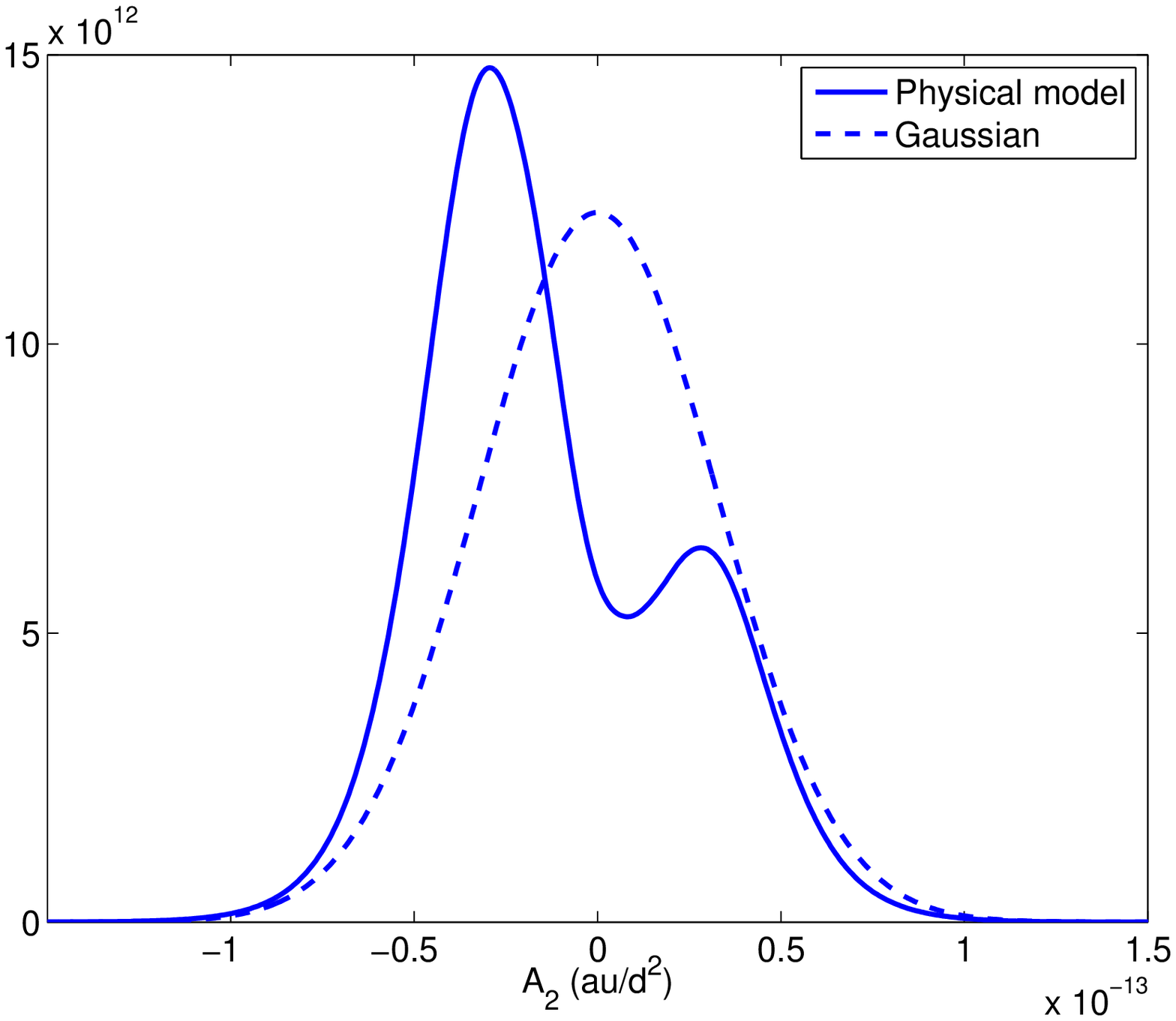}{Distribution of the Yarkovsky parameter
  $A_2$. The solid curve corresponds to the distribution obtained from
  the physical model; the dashed line is a normal distribution with
  the same $3\sigma$ limits.}
\end{figure}

From the described physical model we also obtain a nominal mass of
$8.3 \times 10^{10}$ kg, which we use in Sec.~\ref{sec:apriori}
and \ref{sec:montecarlo} to estimate the energy released by a possible
impact.

If we were to assume that the albedo was $0.06 \pm 0.015$, with
absolute magnitude $H=22.1 \pm 0.3$, then the $3\,\sigma$ uncertainty
would grow to $215.3$. This would imply lower IPs and lower mass
estimates in the results in Sects. \ref{sec:apriori} and
\ref{sec:montecarlo}, but the overall structure of the VIs would be
preserved, possibly with some additional VIs in the distribution
tails.

\section{Line of variations in $>6$ dimensions}
\label{sec:lov}

The most common parameter when modeling the Yarkovsky effect is $A_2$,
i.e., the coefficient appearing in the average transverse
acceleration: $ T={A_2}/{r^2} $, where $r$ is the distance from the
Sun. The result is obtained by fitting the available astrometry
(optical and radar) to the initial conditions and the $A_2$ parameter.
Thus all the orbit determination process has to be done with seven
parameters, the normal matrix $C$ is $7\times 7$, and the eigenvector
$V_1$ of $C$ with smallest eigenvalue is seven-dimensional
\citep[][Chap.~5, 10]{orbdet}. The theory of the line of variations
\citep{multsol} can be generalized to dimension $>6$:
the LOV is defined as the set of the local minima of the target
function restricted to hyperplanes orthogonal to $V_1$. The actual
computation of the LOV uses a constrained differential correction
process operating on this hyperplane. This change is conceptually
straightforward, but in terms of programming it is a
complicated task. As a result, version 4.3 of the software system
OrbFit, implementing a full seven-dimensional LOV and seven-dimensional
impact monitoring, is still undergoing testing and has not yet replaced
the operational version 4.2 \footnote{http://adams.dm.unipi.it/orbfit/}.

However, the impact monitoring processing chain including Yarkovsky
effect has already been tested, in particular on the case of (99942)
Apophis.  The comparison with results obtained with Monte Carlo
method has confirmed that the method gives satisfactory results,
provided one problem is solved.
As discussed in \citep{multsol}, the notion of smallest eigenvalue
depends on a metric in the parameter space, thus it is not invariant
for coordinate changes. For comparatively short term impact monitoring
(a few tens of years) we can select an appropriate coordinate system
depending on the astrometry available (e.g., Cartesian coordinates for
short observed arcs, equinoctal elements \citep{broucke} for longer
arcs).

The best choice of LOV, applicable to a much longer time span, would
need to have the following property.  If there is a planetary
encounter that scatters the LOV solutions into qualitatively different
orbits such that they can result in successive encounters in
different years, then we select the Target Plane (TP) of this
encounter as scattering plane \citep{bennu}.  The best LOV in
the space of initial conditions and parameters is such that the spread
of corresponding TP points is maximum. In this way, all the dynamical
pathways after the scattering encounter, which could lead to succesive
impacts, are represented on the LOV.

To achieve this result, before computing the LOV we propagated the
nominal orbit to the scattering plane, where we found the major axis
vector $W\in \R^2$ of the confidence ellipse obtained by linear
propagation of the orbit covariance. Among the possible inverse images
of $W$ by the differential of the propagation to the TP, we selected
$Z\in \R^7$ corresponding to the minimum increase in the quadratic
approximation to the target function, as given by the appropriate
regression line. We then used $Z$ as the direction of the LOV. For
very well determined orbits such as the one of 2009~FD, given the
direction $Z$, the LOV can be computed as a straight line: a full
nonlinear computation would give negligible changes in the selected
sample points.

\section{Impact monitoring with a priori constraints}
\label{sec:apriori}

We carefully analyzed the available astrometry and manually weighted
the observations to account for the uncertainty information provided
by some of the observers and to mitigate the effect of correlations
for nights with a large number of observations.

When solving for the six orbital elements the orbit is very well
constrained. For instance, the standard deviation for the semimajor
axis $a$ is $STD(a)=1.8\times 10^{-9}$ au $=270$ m. However, if the
seventh parameter $A_2$ is also determined its uncertainty is too
large and the nominal value does not provide useful information. Thus,
we decided to assume an a priori value $A_2=(0\pm 32.5) \times
10^{-15}$ au/d$^2$, consistent with the discussion in
Sec.~\ref{sec:yarko}. The a priori observation was added to the normal
equation with the standard formula \citep[Sec. 6.1]{orbdet}.

In these conditions, the best fit value is $A_2=(-2 \pm 32.5) \times
10^{-15}$au/d$^2$, which is not significantly different from $0$. The
$STD(a)=2.3\times 10^{-9}$ au is not much higher that the
six-parameter fit. We then run the computation of the LOV defined by
the 2185 scattering plane, with $2\, 401$ points up to $|\sigma|=3$,
the propagation to the year 2250 of all the sample points on the LOV,
and the search for virtual impactors, all in the seven-dimensional
version; apart from the change in dimension, the method is not
different from \citep{clomon2}.

These computations were done with DE430 planetary ephemerides
from JPL \citep{de430}, 17 perturbing asteroids including Pluto, and
appropriate relativistic dynamics as discussed in \citep{bennu}. To
assess the risk level, we computed the Palermo Scale (PS) by
using the expected value of the mass as estimated in
Sec.~\ref{sec:yarko}.

\begin{table}[ht]
  \footnotesize
  \centering
  \caption{Risk file for 2009~FD: calendar year, month, and day for the
    potential impact; approximate location along the LOV in $\sigma$
    space; minimum distance (the lateral distance from the LOV to the
    center of the Earth with the 1 $\sigma$ semi-width of the TP
    confidence region); stretching factor (how much the confidence
    region at the epoch has been stretched by the time of impact);
    probability of Earth impact; and Palermo Scale. The width of the
    TP confidence region is always few km. For all VIs the LOV
    directly intersects the Earth. }
  \label{tab:riskNEODyS}
  \medskip
  \begin{tabular}{clcrcc}
    \hline
    date       &          sigma & dist& stretch& IP$_{\oplus}$        & PS\\
 yyyy-mm-dd .dd &                & (r$_{\oplus}$)&        &                    &   \\
    \hline
2185-03-29.75 &          $-1.069$ &0.52 &     184& 2.71 $\times 10^{-3}$&$-0.43$ \\
2186-03-29.98 &          $-1.049$ &0.58 & 1450000& 3.50 $\times 10^{-7}$&$-4.32$ \\
2190-03-30.08 &$\phantom{-}0.005$ &0.57 &    2960& 2.92 $\times 10^{-4}$&$-1.41$ \\
2191-03-30.21 &          $-0.962$ &0.89 &  377000& 1.24 $\times 10^{-6}$&$-3.78$ \\
2192-03-29.51 &          $-1.003$ &0.87 & 1110000& 3.96 $\times 10^{-7}$&$-4.28$ \\
2194-03-30.02 &          $-1.025$ &0.93 & 3110000& 1.58 $\times 10^{-7}$&$-4.68$ \\
2196-03-29.44 &          $-0.872$ &0.54 &  225000& 2.68 $\times 10^{-6}$&$-3.46$ \\
    \hline
  \end{tabular}
\end{table}

Table~\ref{tab:riskNEODyS} includes the main 2185 VI with the highest
$PS=-0.43$ among all asteroids currently on our Risk Page. Its
$IP\simeq 1/369$ is quite high, especially for an impact with an
estimated energy of $\simeq 3\,700$ Mt of TNT. On the contrary, the IP
in 2190 is lower than that computed with a purely gravitational model,
although the current VI is very close to the nominal solution. The
computations with the Yarkovsky effect were crucial for a reliable
assessment of the impact risk.

\begin{figure}[t]
  \figfigincl{12 cm}{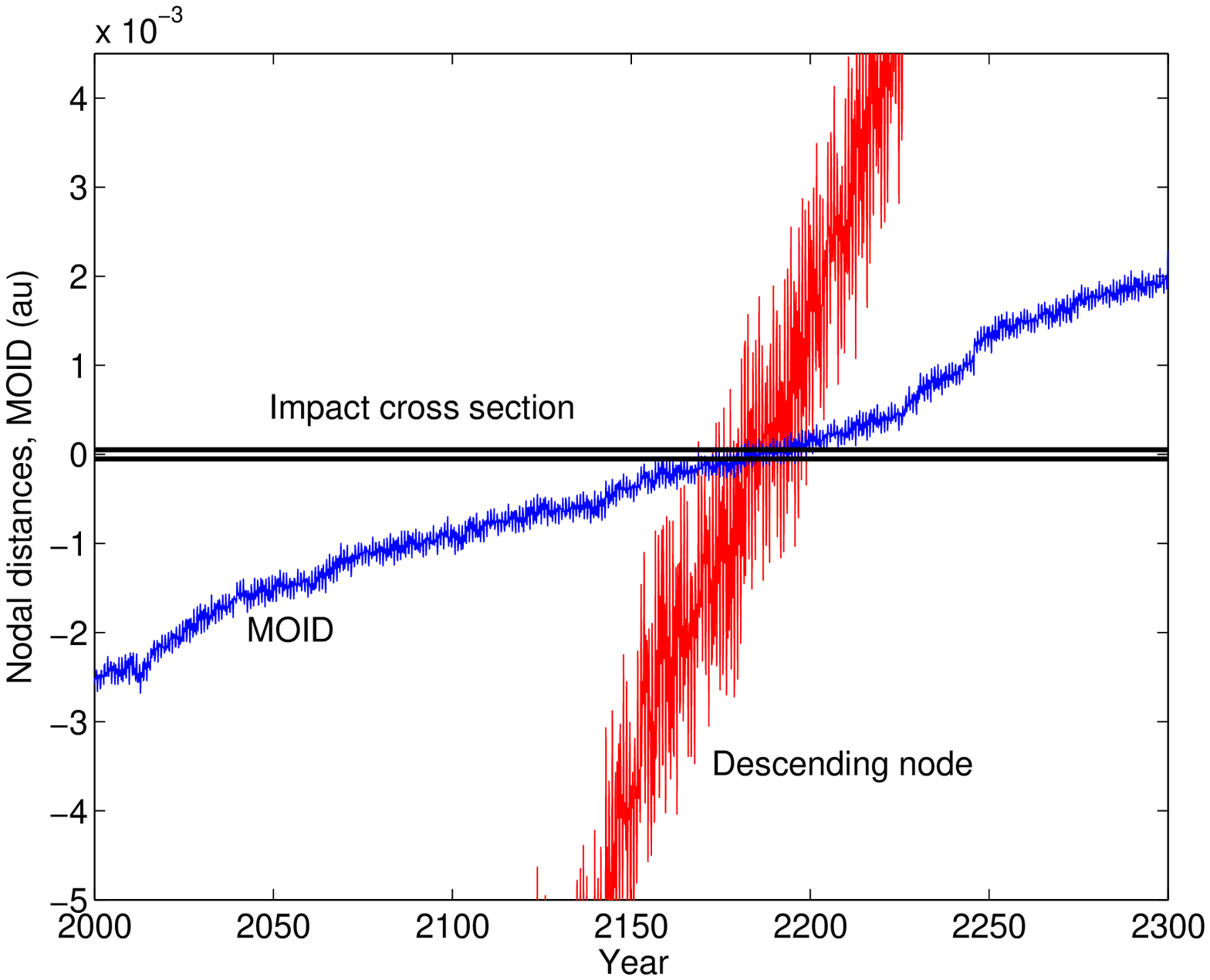}{From the orbit of 2009~FD
    propagated until year 2300, we have computed the MOID and the
    distance at the descending node (in au). A MOID smaller than the
    radius of the impact cross-section occurs between 2166 and
    2197.}
\end{figure}

We performed the impact monitoring with limit date in 2250 and we
found many close approaches in every single year until 2250, but none
leading to impact because of the secular increase in the
Minimum Orbit Intersection Distance (MOID) (see
Fig.~\ref{fig:2009FD_moid}); the closest one is in 2198 with a close
approach of $1.4$ Earth radii.

\section{Monte Carlo impact monitoring}
\label{sec:montecarlo}

The JPL Sentry risk assessment was independently performed by means of
a Monte Carlo simulation. First, we computed a seven-dimensional
orbital solution, with $A_2 = 0$ au/d$^2$. The a priori uncertainty on
$A_2$ was set to obtain a postfit uncertainty of $32.5$ au/d$^2$.
Then, we used the resulting seven-dimensional covariance to randomly
generate a million samples thus getting a resolution of $\sim 10^{-6}$
for the impact probability. Finally, we propagated each sample,
recorded the future close approaches, and counted the impacts
occurring before 2200. The dynamical model is the same used for the
computations in Sec.~\ref{sec:apriori}.

Table~\ref{tab:riskSentry} lists the possible impacts found by the
Monte Carlo method. The values of $\sigma$ are computed by taking the
distribution of the Monte Carlo samples on the 2185 $b$-plane.  The
impact probabilities do not change very much if we use the most
complete information about the A2 distribution in the Monte Carlo
method. The ratio between the A2 distribution from the physical model
and the normal A2 distribution is about $2$, except the left tails of
the distribution.  However, the impacts found in the Monte Carlo
simulations are for $|A2| < 5 \times 10^{-15}$ au/d$^2$ and are
therefore far from the distribution tails.
\begin{table}[ht]
  \footnotesize
  \centering
  \caption{The JPL Sentry risk file for 2009~FD obtained from the
    Monte Carlo simulation: date for the potential impact; approximate
    location along the LOV in sigma space; minimum distance;
    stretching factor; probability of Earth impact; and Palermo Scale.}
  \label{tab:riskSentry}
  \medskip
  \begin{tabular}{clcrcc}
    \hline
    date       & sigma           & dist& stretch  & IP$_{\oplus}$   & PS\\
 yyyy-mm-dd.dd &                 & (r$_{\oplus}$)&          &               &   \\
    \hline
2185-03-29.75 &           $-1.048$ & 0.52 &    160& 2.6 $\times 10^{-3}$& $-0.44$\\
2190-03-30.08 & $\phantom{-}0.028$ & 0.57 &   2580& 2.7 $\times 10^{-4}$& $-1.43$\\
2191-03-30.24 &           $-0.993$ & 0.40 & 209000& 2.0 $\times 10^{-6}$& $-3.57$\\
2192-03-29.51 &           $-0.986$ & 0.83 & 364000& 1.0 $\times 10^{-6}$& $-3.87$\\
2196-03-29.44 &           $-0.864$ & 0.76 & 496000& 1.0 $\times 10^{-6}$& $-3.88$\\
    \hline
  \end{tabular}
\end{table}

\section{Scattering encounter}
\label{sec:scattering}

\begin{figure}[h]
\figfigincl{12 cm}{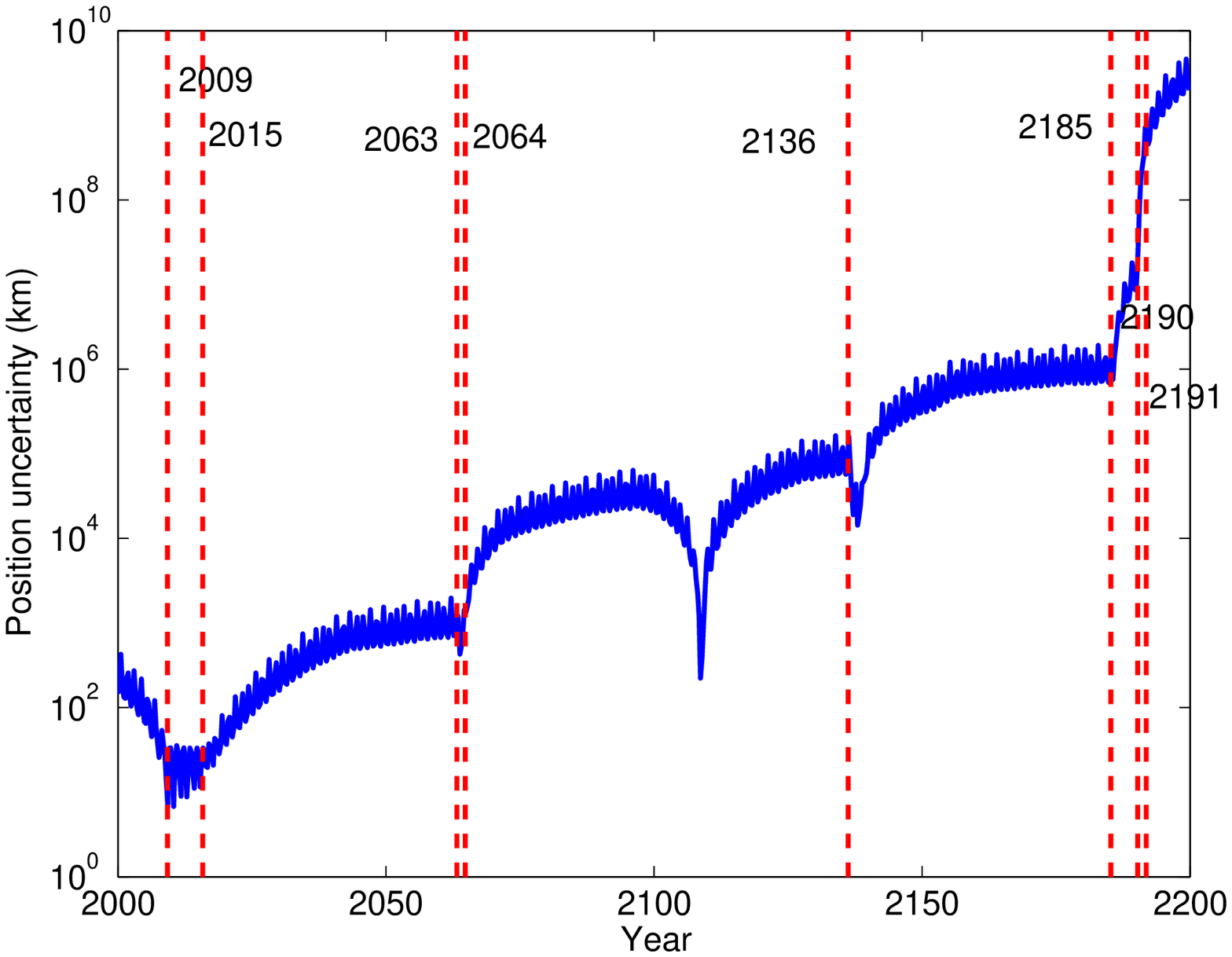}{Evolution of the longest semiaxis of
  the $1\,\sigma$ confidence ellipsoid. The vertical dashed lines
  correspond to close approaches with the Earth within $0.05$ au.}
\end{figure}

Figure~\ref{fig:unc_2009fd} shows the increase of the position
uncertainty (longest semiaxis of the $1\,\sigma$ confidence ellipsoid,
as deduced from the linearly propagated confidence matrix) with
time. This increase is by no means a gradual increase, but mostly
occurs within short time intervals corresponding to close approaches
to the Earth. In particular, the position uncertainty increases as
$\Delta t^2$ far from the close approaches, while it increases much
more quickly during the close approaches. The deepest close approaches
are marked by dotted lines. After the close approach in 2009, there is
another at a minimum distance of $0.418$ au in 2015, then the pair in
2063 (at $0.0130$ au) and 2064 (at $0.0266$ au). The last two are near
different nodes, connected by a nonresonant return
\citep{1999AN10}. Another nonresonant return connects the close
approach of 2136 (at $0.0218$ au) with the shallower one of 2137 (at
$0.0815$ au). In all cases, the divergence of nearby orbits
fluctuates, but overall increases by a factor of $\simeq 3 \times
10^5$ over 176 years (from 2009 to 2185), corresponding to a Lyapounov
time $176/\log(3 \times 10^5)\simeq 14$ years. Over this timespan the
Yarkovsky effect is very important, because the prediction uncertainty
is not dominated by the chaotic effects.

\begin{figure}
\centering
\begin{minipage}{.5\textwidth}
\figfiginclnocap{6 cm}{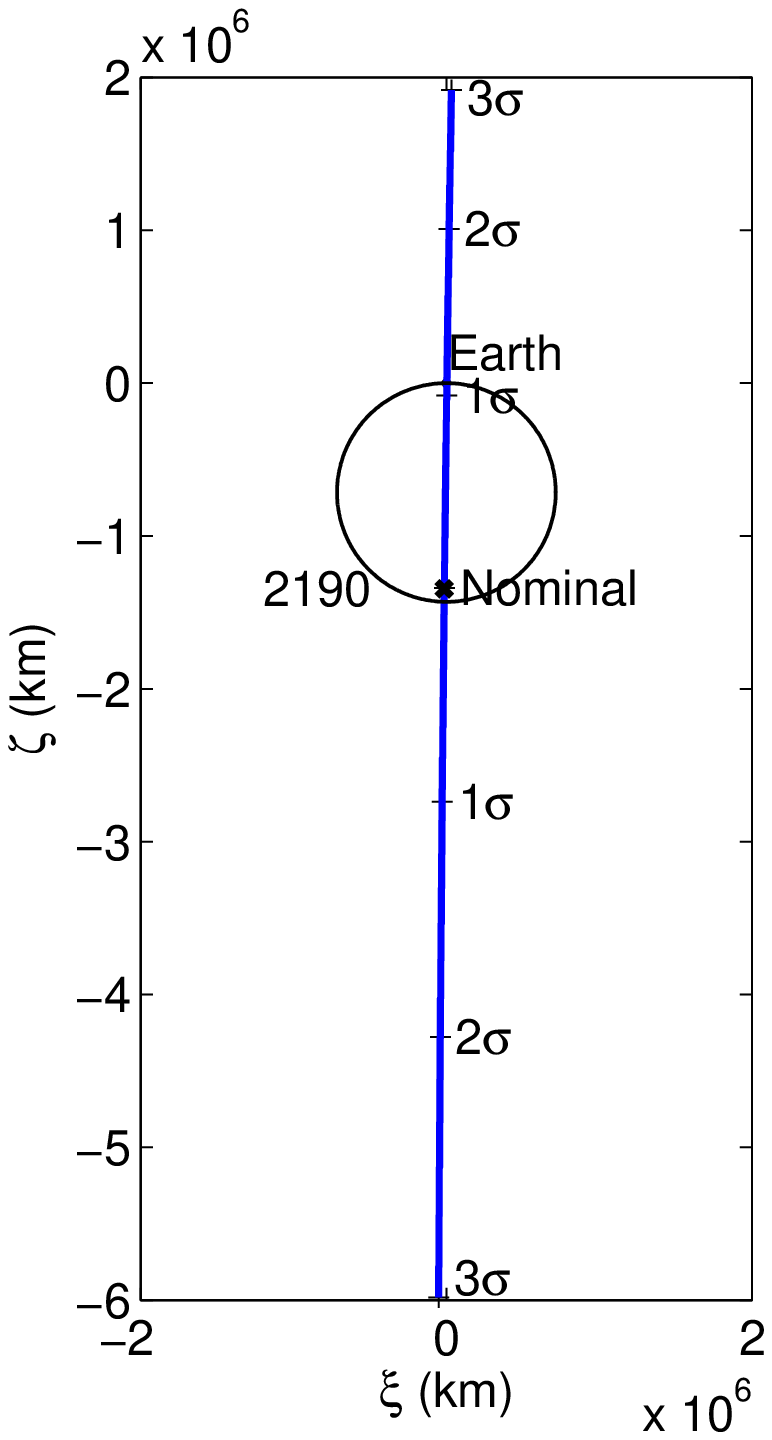}
\end{minipage}%
\begin{minipage}{.5\textwidth}
 \figfiginclnocap{6 cm}{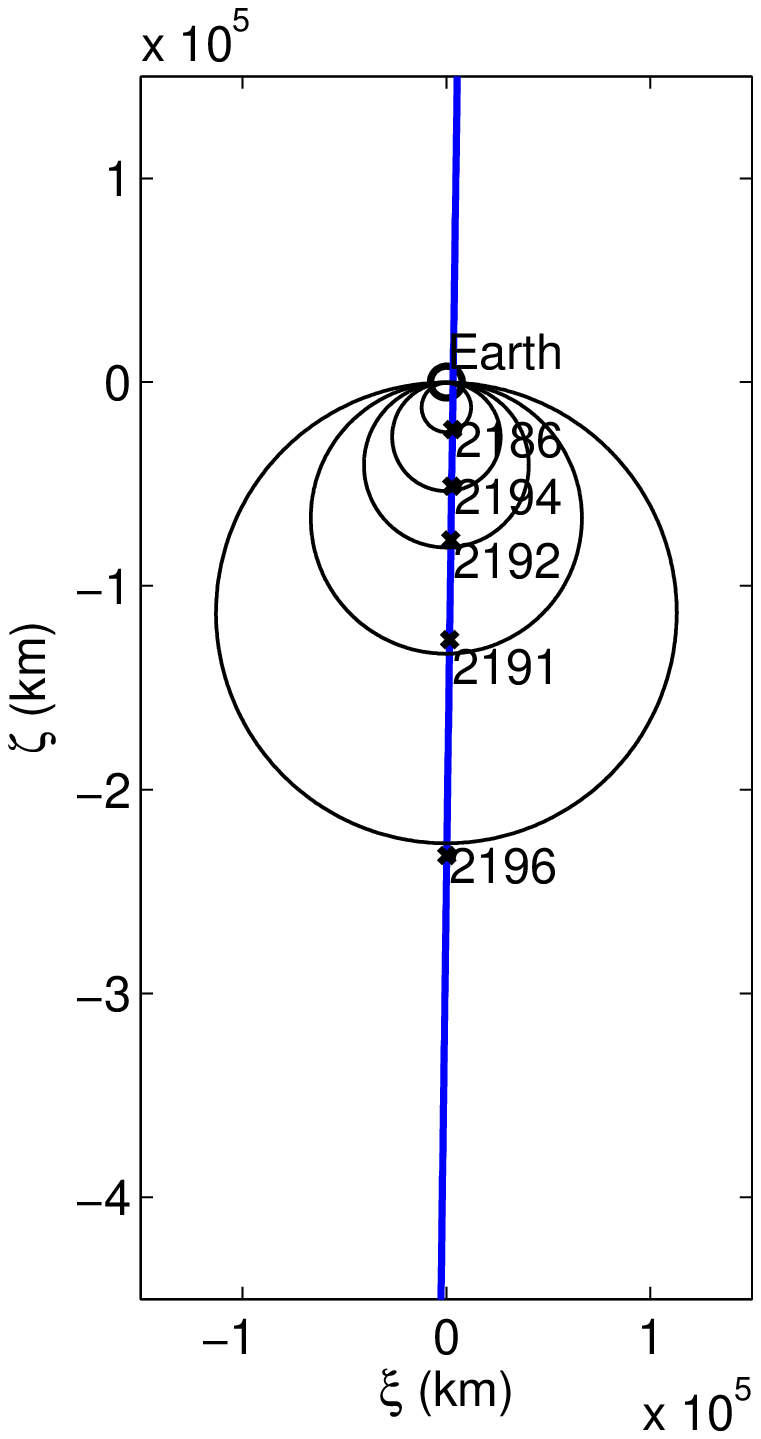}
\end{minipage}
\caption{The left figure shows the LOV on the 2185 TP: the VI of 2190
  is marked with a cross.  The right figure shows a segment of the LOV
  on the 2185 TP: the keyholes for impact in (from top to bottom)
  2186, 2194, 2192, 2191, and 2196 are marked with crosses; also shown
  are the $b$-plane circles associated with the respective mean motion
  resonances (1/1, 8/9 6/7, 5/6, 9/11).}\label{fig:circles}
\end{figure}

Finally, as Fig. \ref{fig:circles} shows, at the time of the 2185
encounter the LOV (plotted over the interval from $\sigma=-3$ to $+3$)
spans more than 7 million km on the $b$-plane, and straddles the
Earth.  As a consequence, a very wide range of close approach
distances is possible, from actual Earth collision up to
very distant encounters. The 2185 VI is similar to a direct impact,
with a very low stretching, hence the comparatively large IP.

After 2185, the divergence grows to much larger values, of course
different for the different LOV sample orbits, depending on how
close the 2185 encounter is, and the prediction uncertainty becomes
dominated by chaos. The later VI detected have higher stretching, thus
lower IP, and are resonant returns \citep{1999AN10} from comparatively
close approaches in 2185, occurring at the same date after 1, 5, 6, 7,
11 years. They correspond to resonances between the mean motions
$n,n_\oplus$ of the asteroid and the Earth, respectively:
\[
\frac{n}{n_\oplus}=\frac 11 \;,\; \frac 45 \;,\; \frac 56 \;,\; \frac
67 \;,\; \frac 89 \;,\; \frac 9{11}\ \ .
\]

Figure~\ref{fig:circles} shows the location of the VI on the
LOV and the Valsecchi circles corresponding to the resonances,
according to an approximate analytic theory \citep{analytic}.  The
$4/5$ resonance corresponds to the weakest perturbation in 2185
since the orbit is currently close to it: thus the circle for returns
in 2190, shown in the left plot of Fig.~\ref{fig:circles}, is much larger
than the others, shown at a larger scale in the right plot.

It can be shown with formulas derived from \citep{analytic} that the
values of the semimajor axis after the 2185 encounter could range
between $0.82$ and $2.10$ au, these values being obtained for grazing
encounters (see Appendix~\ref{sec:appendix}). This semimajor axis gives
access to all the resonances with ratio of mean motions ranging from
$4/3$ to $1/3$. In the propagation of the LOV orbits we find
close approaches to the Earth occurring in each single year after 2185
until the growth of the MOID prevents the possibility of impact after
2196.

Thus the analytic theory allows us to predict the approximate location
of possible keyholes \citep{chodas}.
Figure~\ref{fig:keyholes_map_zoom} compares the analytical and the
numerical computations of locations and widths of the keyholes, from
which the IPs can be computed. However, the analytical theory does not
take into account the short periodic changes in the MOID (see
Fig.~\ref{fig:2009FD_moid}), thus it can only provide upper bounds for
the widths and the IPs: it is expected that the number of potential
keyholes computed analytically will be greater than the actual
keyholes found numerically. It is also interesting to compare the
probability density distributions computed with or without the
Yarkovky effect. As a result of this comparison, the $3\, \sigma$
value for the distribution obtained with the Yarkovsky effect is
$\sim10^6$ km, while the same value for the distribution without the
Yarkovky effect is $\sim 10^5$ km.

\begin{figure}[h]
\figfiginclnocap{12 cm}{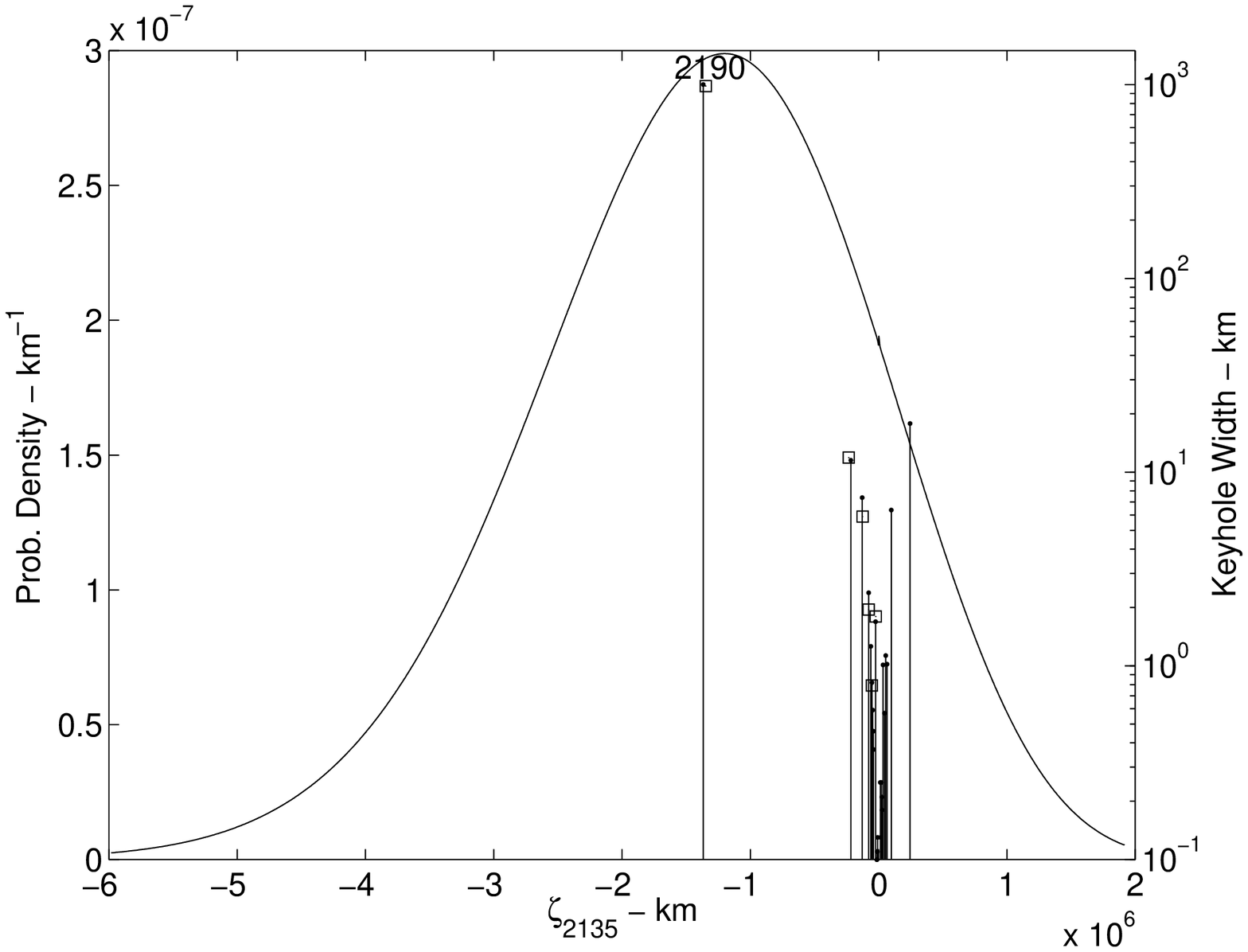}
\end{figure}

\begin{figure}[h]
  \figfigincl{12 cm}{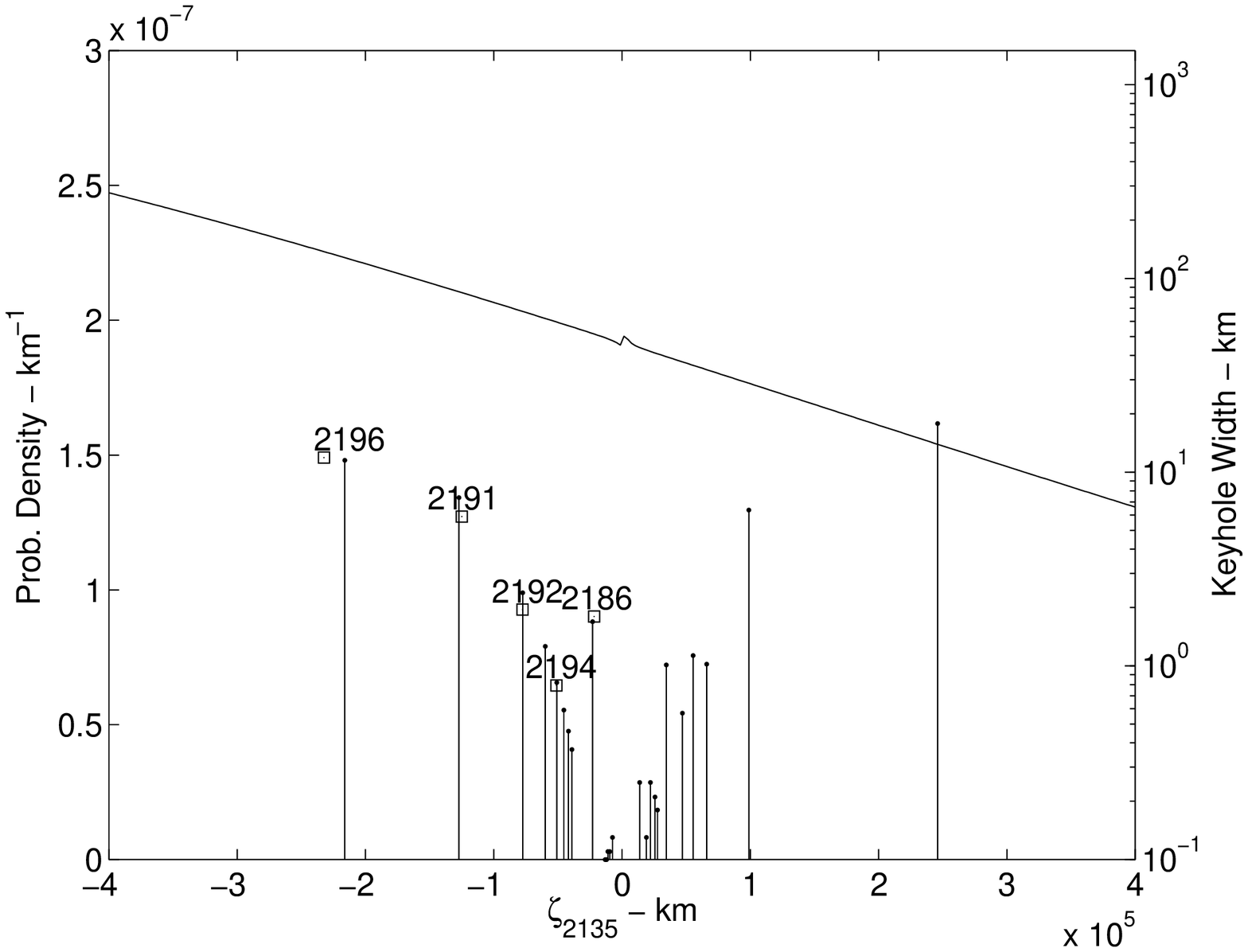}{Map of the 2009~FD impact
    keyholes intersecting the trace of the LOV on the 2185 TP,
    computed both numerically and analitically. The probability
    density is given by the curve with the left scale, and the
    analytically computed keyholes are indicated by the vertical lines
    with their widths given by the height of the bar (right
    scale). The 7 actual VIs found numerically (from
    Table~\ref{tab:riskNEODyS}) are marked with a square.}
\end{figure}

\section{Conclusions}
\label{sec:compare}

\subsection{Comparison and reliability of the results}
 
Given the use of the same dynamical model, it is no surprise that the
results are very similar: still, they are remarkably close.  The LOV
method finds seven VIs, while the Monte Carlo method finds five, but
the two missing VIs have IPs of $3.5\times 10^{-7}$ and $1.6\times
10^{-7}$, below the sensitivity of the Monte Carlo method with a
million samples. The five common VIs have consistent IP estimates: the
higher ones, for 2185 and 2190, are in agreement within a few
percentage points; the three lower ones are different (as expected)
because of Poisson statistics.

Thus the Monte Carlo method detects VI down to its sensitivity limit,
which is $1\times 10^{-6}$; the LOV method detects VI close to its
generic completeness limit \citep{clomon2}[Sec. 2.5], which is
$IP\simeq 2.5\times 10^{-7}$. We note that it would not be difficult
to upgrade these sensitivity limits, with both methods, by increasing
the resolution and therefore the computational load.  The seven
dimensional LOV method is more efficient from the computational point
of view for a given resolution. On the other hand, the Monte Carlo
method is simpler from the software perspective and is generally more
reliable, e.g., in the case of off--LOV VIs.

Thus, the problem we had with the computation of the possibilities of
impact and the values of the IP was solved, and we posted the new
results on our online services NEODyS and Sentry.

The results obtained with the analytical theory despite the
approximations it must contain are remarkably good, to the point that
they can be used as a check of the numerical results. However, they
cannot replace numerical computations to verify that some VIs actually
exists in an accurate orbital computation.

\subsection{Future observability}

2009 FD could be optically observable again during its next
apparition in early 2015. 2009~FD should become detectable with a
large-aperture telescope (8-meter class) in October 2014, and even with
smaller apertures (2-meter class) around January 2015, when it is expected
to reach a peak magntiude of $V=23$ near opposition. However, the very
small skyplane uncertainty of this apparition ($0.15''$ at the
$3\sigma$ level, even including the contribution due to Yarkovsky)
will prevent any significant improvement in the overall orbital
uncertainty.

The next valuable opportunity to collect useful information will begin
in late October 2015, when 2009~FD will emerge from solar conjunction
immediately after its closest approach, already at $V=19$.  The
magnitude will then reach a peak magnitude of $V=18$ within a few
days, making 2009~FD an easy target for physical observations from the
ground even with modest apertures. On 2015 November 1 the $3\,\sigma$
uncertainty ellipse will have semiaxes 2.34'' and 0.69'' (most
uncertainty is in declination, while proper motion is mostly along
right ascension). This uncertainty will give significant leverage for
orbital improvement with ground-based astrometry.

Since the late October 2015 close approach will be at about $0.04$ au
from the geocenter, there could be radar observations, hopefully
including range measurements (which were not possible in 2014, with
minimum distance $0.1$ au).

If these radar and optical observations improve the constraints on the
Yarkovsky parameter $A_2$, then the Impact Monitoring will need to be
recomputed. Our methods and software now allow us to do this, although
a manual procedure is still required: it involves a limited amount of
manpower and some computing time.

The object will then slowly approach solar conjunction during the
first half of 2016, and will not become easily observable from the
ground until late 2018 (apart from a couple of very challenging
low-elongation opportunities in early and late 2017 of very limited
astrometric relevance). Unless the 2015 observations rule out the two
dangerous segments of the current LOV, observations may need to be
continued for a long time, before the 2009~FD impact problem is
resolved.

\section*{Acknowledgments}

The work of F. Spoto and A. Milani has been supported by the
Department of Mathematics, University of Pisa, with an internal
grant. The work of D. Farnocchia and S.R. Chesley was conducted at the
Jet Propulsion Laboratory, California Institute of Technology under a
contract with the National Aeronautics and Space Administration.  The
work of M. Micheli was funded under ESA contract
No. 4000107291/13/D/MRP–-SSA NEO Segment Precursor Service Operations
(SN-V). D. Perna acknowledges financial support from the NEOShield
project, funded by the European Commission's Seventh Framework
Programme (Contract No. FP7-SPACE-2011-282703).

\bibliographystyle{aa}
\bibliography{biblio}

\begin{thebibliography}{21}
\expandafter\ifx\csname natexlab\endcsname\relax\def\natexlab#1{#1}\fi

\bibitem[{{Broucke} \& {Cefola}(1972)}]{broucke}
{Broucke}, R.~A. \& {Cefola}, P.~J. 1972, Celestial Mechanics, 5, 303

\bibitem[{{Carbognani}(2011)}]{rotP}
{Carbognani}, A. 2011, Minor Planet Bulletin, 38, 57

\bibitem[{{Chesley} {et~al.}(2014){Chesley}, {Farnocchia}, {Nolan},
  {Vokrouhlick{\'y}}, {Chodas}, {Milani}, {Spoto}, {Rozitis}, {Benner},
  {Bottke}, {Busch}, {Emery}, {Howell}, {Lauretta}, {Margot}, \&
  {Taylor}}]{bennu}
{Chesley}, S.~R., {Farnocchia}, D., {Nolan}, M.~C., {et~al.} 2014, Icarus, 235,
  5

\bibitem[{{Chodas}(1999)}]{chodas}
{Chodas}, P.~W. 1999, in Bulletin of the American Astronomical Society,
  Vol.~31, Bulletin of the American Astronomical Society, 1117

\bibitem[{{Delb\`o} {et~al.}(2007){Delb\`o}, {dell'Oro}, {Harris}, {Mottola},
  \& {Mueller}}]{thermal_inertia}
{Delb\`o}, M., {dell'Oro}, A., {Harris}, A.~W., {Mottola}, S., \& {Mueller}, M.
  2007, Icarus, 190, 236

\bibitem[{{DeMeo} {et~al.}(2009){DeMeo}, {Binzel}, {Slivan}, \& {Bus}}]{demeo}
{DeMeo}, F.~E., {Binzel}, R.~P., {Slivan}, S.~M., \& {Bus}, S.~J. 2009, Icarus,
  202, 160

\bibitem[{{Farnocchia} \& {Chesley}(2014)}]{1950DA}
{Farnocchia}, D. \& {Chesley}, S.~R. 2014, Icarus, 229, 321

\bibitem[{{Farnocchia} {et~al.}(2013{\natexlab{a}}){Farnocchia}, {Chesley},
  {Chodas}, {Micheli}, {Tholen}, {Milani}, {Elliott}, \& {Bernardi}}]{apophis}
{Farnocchia}, D., {Chesley}, S.~R., {Chodas}, P.~W., {et~al.}
  2013{\natexlab{a}}, Icarus, 224, 192

\bibitem[{{Farnocchia} {et~al.}(2013{\natexlab{b}}){Farnocchia}, {Chesley},
  {Vokrouhlick{\'y}}, {Milani}, {Spoto}, \& {Bottke}}]{yarko_list}
{Farnocchia}, D., {Chesley}, S.~R., {Vokrouhlick{\'y}}, D., {et~al.}
  2013{\natexlab{b}}, Icarus, 224, 1

\bibitem[{{Folkner} {et~al.}(2014){Folkner}, {Williams}, {Boggs}, {Park}, \&
  {Kuchynka}}]{de430}
{Folkner}, W.~M., {Williams}, J.~G., {Boggs}, D.~H., {Park}, R.~S., \&
  {Kuchynka}, P. 2014, Interplanetary Network Progress Report, 196, C1

\bibitem[{{Gwyn} {et~al.}(2012){Gwyn}, {Hill}, \& {Kavelaars}}]{gwyn}
{Gwyn}, S.~D.~J., {Hill}, N., \& {Kavelaars}, J.~J. 2012, Astronomical Data
  Analysis Software and Systems XXI, 124, 579

\bibitem[{{La Spina} {et~al.}(2004){La Spina}, {Paolicchi}, {Kryszczy{\'n}ska},
  \& {Pravec}}]{laspina}
{La Spina}, A., {Paolicchi}, P., {Kryszczy{\'n}ska}, A., \& {Pravec}, P. 2004,
  Nature, 428, 400

\bibitem[{{Landolt}(1992)}]{landolt}
{Landolt}, A.~U. 1992, Astronomical Journal, 104, 340

\bibitem[{{Mainzer} {et~al.}(2014){Mainzer}, {Bauer}, {Grav}, {Masiero},
  {Cutri}, {Wright}, {Nugent}, {Stevenson}, {Clyne}, {Cukrov}, \&
  {Masci}}]{wise}
{Mainzer}, A., {Bauer}, J., {Grav}, T., {et~al.} 2014, The Astrophysical
  Journal, 784, 110

\bibitem[{{Milani} {et~al.}(2005{\natexlab{a}}){Milani}, {Chesley},
  {Sansaturio}, {Tommei}, \& {Valsecchi}}]{clomon2}
{Milani}, A., {Chesley}, S.~R., {Sansaturio}, M.~E., {Tommei}, G., \&
  {Valsecchi}, G.~B. 2005{\natexlab{a}}, Icarus, 173, 362

\bibitem[{{Milani} {et~al.}(1999){Milani}, {Chesley}, \&
  {Valsecchi}}]{1999AN10}
{Milani}, A., {Chesley}, S.~R., \& {Valsecchi}, G.~B. 1999, Astronomy and
  Astrophysics, 346, L65

\bibitem[{{Milani} \& {Gronchi}(2010)}]{orbdet}
{Milani}, A. \& {Gronchi}, G.~F. 2010, {Theory of Orbital Determination}
  (Cambridge University Press)

\bibitem[{{Milani} {et~al.}(2005{\natexlab{b}}){Milani}, {Sansaturio},
  {Tommei}, {Arratia}, \& {Chesley}}]{multsol}
{Milani}, A., {Sansaturio}, M.~E., {Tommei}, G., {Arratia}, O., \& {Chesley},
  S.~R. 2005{\natexlab{b}}, Astronomy and Astrophysics, 431, 729

\bibitem[{{Mommert} {et~al.}(2014){Mommert}, {Hora}, {Farnocchia}, {Chesley},
  {Vokrouhlick{\'y}}, {Trilling}, {Mueller}, {Harris}, {Smith}, \&
  {Fazio}}]{2009bd}
{Mommert}, M., {Hora}, J.~L., {Farnocchia}, D., {et~al.} 2014, The
  Astrophysical Journal, 786, 148

\bibitem[{{Valsecchi} {et~al.}(2003){Valsecchi}, {Milani}, {Gronchi}, \&
  {Chesley}}]{analytic}
{Valsecchi}, G.~B., {Milani}, A., {Gronchi}, G.~F., \& {Chesley}, S.~R. 2003,
  The Astrophysical Journal, 408, 1179

\bibitem[{{Vokrouhlick{\'y}} {et~al.}(2000){Vokrouhlick{\'y}}, {Milani}, \&
  {Chesley}}]{yarkopred}
{Vokrouhlick{\'y}}, D., {Milani}, A., \& {Chesley}, S.~R. 2000, Icarus, 148,
  118

\end{thebibliography}

\begin{appendix}
\section{An analytic estimate of the resonant returns cascade}
\label{sec:appendix}

We can make an analytic estimate of the range of semimajor axes of the
possible post-2185 orbits \citep{analytic}; in doing so, we will use
the $b$-plane coordinates $\xi$, which correspond to the local MOID
with sign, and $\zeta$, which is related to the timing of the encounter, as
well as the values of the unperturbed geocentric velocity $U$ (in
units of the Earth orbital velocity), and of the angle $\theta$
between the velocity of the Earth and the unperturbed geocentric
velocity of 2009 FD at the encounter of 2185.  We assume the values of
the 2185 VI, $U=0.533$ and $\theta=97^\circ\!\!.7$.

We then compute $c=m_\oplus/U^2$, where $m_\oplus$ is the mass of the
Earth in units of the solar mass; $c$ is the value of the impact
parameter leading to a rotation of the geocentric velocity by
$90^\circ$, and plays the role of a characteristic length for each
NEA.  In the case of 2009~FD $c=0.25$ $r_\oplus$, where $r_\oplus$ is
the Earth radius.

The gravitational cross section of the Earth seen by 2009~FD is a disk
of radius $b_\oplus$ \citep{analytic},
\begin{displaymath}
b_\oplus=r_\oplus\sqrt{1+\frac{2c}{r_\oplus}}=1.22\,r_\oplus;
\end{displaymath}
thus, the $b$-plane distance corresponding to a grazing Earth
encounter is $1.22\,r_\oplus$.

We now turn to the possible post-encounter values of the orbital
semimajor axis $a'$ of 2009~FD, which is given by
\begin{displaymath}
a'=\frac{1}{1-U^2-2U\cos\theta'};
\end{displaymath}
in fact, as discussed by \citep{analytic}, $a'$ is maximum when
$\cos\theta'$ is maximum, and $a'$ is minimum when $\cos\theta'$ is
minimum.  We can therefore consider the expression for $\cos\theta'$
as a function of the $b$-plane coordinates:
\begin{displaymath}
\cos\theta'=\frac{(\xi^2+\zeta^2-c^2)\cos\theta+2c\zeta\sin\theta}{\xi^2+\zeta^2+c^2}.
\end{displaymath}
We use the wire approximation of \citep{analytic}, so that $\xi$
can be considered constant, like all other quantities in the
expression, except $\zeta$. We therefore take the partial derivative
with respect to $\zeta$,
\begin{eqnarray*}
\frac{\partial\cos\theta'}{\partial\zeta} 
& = & \frac{2c[2c\zeta\cos\theta+(\xi^2-\zeta^2+c^2)\sin\theta]}{(\xi^2+\zeta^2+c^2)^2},
\end{eqnarray*}
and look for the zeroes $\zeta_\pm$ of its numerator:
\begin{eqnarray*}
0 & = & \zeta^2\sin\theta-2c\zeta\cos\theta-(\xi^2+c^2)\sin\theta \\
\zeta_\pm 
& = & \frac{c\cos\theta\pm\sqrt{c^2+\xi^2\sin^2\theta}}{\sin\theta}.
\end{eqnarray*}
Making the appropriate substitutions ($c=0.25\,r_\oplus,
|\xi|=0.52\,r_\oplus, \theta=97^\circ\!\!.7$), we get
$\zeta_+=0.54\,r_\oplus$ and $\zeta_-=-0.61\,r_\oplus$; both values
lead to values smaller than $b_\oplus$, implying that the maximum and
minimum possible values for $a'$ are obtained for grazing encounters
taking place at $\zeta=\pm\sqrt{b^2_\oplus-\xi^2}=\pm1.11\,r_\oplus$.
Thus, the maximum post-encounter $a'$ and the related maximum orbital
period $P'$ are
\[
a'_{max}  =  2.10~\mathrm{au} \ \ \ and \ \ \
P'_{max}  =  3.05~\mathrm{yr},
\]
and the minimum post-encounter $a'$ and the related minimum orbital
period $P'$ are
\[
a'_{min} =  0.82~\mathrm{au} \ \ \ and \ \ \ 
P'_{min}  =  0.74~\mathrm{yr}.
\]
This range of post-2185 orbital periods for 2009~FD makes a
number resonant of returns within 2196 possible, the year after which the
secular increase in the MOID precludes the possibility of additional
collisions with the Earth at the same node.  The list of resonances is
given in Table~\ref{resonances}; the lines in boldface describe cases
in which actual VIs are found numerically.

\begin{table}[h]
\caption{The resonances with the mean motion of the Earth made
  accessible to 2009~FD by the 2185 close encounter. The lines in boldface show the resonances for
  which actual VIs are found numerically.}
\label{resonances}
\begin{tabular}{c|r@{/}lrrr|r}
\hline
year & \multicolumn{2}{c}{reson.} & $a'$ (au) & $\zeta$ (km) 
& \multicolumn{1}{c}{$\partial\zeta''/\partial\zeta$} & \multicolumn{1}{c}{$P_{max}$} \\
\hline
2194 & 7 &  9 & 1.1824 & 246116 & $8.8\times10^2$ & $2.7\times10^{-6}$ \\
2189 & 3 &  4 & 1.2114 &  98818 & $2.5\times10^3$ & $1.1\times10^{-6}$ \\
2196 & 8 & 11 & 1.2365 &  66047 & $1.5\times10^4$ & $1.9\times10^{-7}$ \\
2192 & 5 &  7 & 1.2515 &  55426 & $1.4\times10^4$ & $2.1\times10^{-7}$ \\
2195 & 7 & 10 & 1.2684 &  47028 & $2.8\times10^4$ & $1.0\times10^{-7}$ \\
2188 & 2 &  3 & 1.3104 &  34570 & $1.5\times10^4$ & $1.9\times10^{-7}$ \\
2196 & 7 & 11 & 1.3517 &  27698 & $8.9\times10^4$ & $3.3\times10^{-8}$ \\
2193 & 5 &  8 & 1.3680 &  25736 & $7.5\times10^4$ & $3.9\times10^{-8}$ \\
2190 & 3 &  5 & 1.4057 &  22199 & $6.3\times10^4$ & $4.7\times10^{-8}$ \\
2197 & 7 & 12 & 1.4324 &  20288 & $1.8\times10^5$ & $1.6\times10^{-8}$ \\
2192 & 4 &  7 & 1.4522 &  19089 & $1.2\times10^5$ & $2.5\times10^{-8}$ \\
2194 & 5 &  9 & 1.4797 &  17667 & $1.8\times10^5$ & $1.7\times10^{-8}$ \\
2196 & 6 & 11 & 1.4979 &  16850 & $2.4\times10^5$ & $1.2\times10^{-8}$ \\
2187 & 1 &  2 & 1.5874 &  13820 & $6.3\times10^4$ & $4.7\times10^{-8}$ \\
2196 & 5 & 11 & 1.6915 &  11524 & $4.8\times10^5$ & $6.2\times10^{-9}$ \\
2194 & 4 &  9 & 1.7171 &  11083 & $4.2\times10^5$ & $7.0\times10^{-9}$ \\
2192 & 3 &  7 & 1.7592 &  10431 & $3.6\times10^5$ & $8.2\times10^{-9}$ \\
2197 & 5 & 12 & 1.7926 &   9970 & $6.7\times10^5$ & $4.4\times10^{-9}$ \\
2190 & 2 &  5 & 1.8420 &   9361 & $3.1\times10^5$ & $9.6\times10^{-9}$ \\
2193 & 3 &  8 & 1.9230 &   8511 & $5.7\times10^5$ & $5.3\times10^{-9}$ \\
2196 & 4 & 11 & 1.9629 &   8146 & $8.3\times10^5$ & $3.6\times10^{-9}$ \\
2188 & 1 &  3 & 2.0801 &   7220 & $2.6\times10^5$ & $1.2\times10^{-8}$ \\
2188 &  4 &  3 & 0.8255 &  $-7389$ & $1.2\times10^5$ & $2.5\times10^{-8}$ \\
2192 &  9 &  7 & 0.8457 &  $-8493$ & $2.6\times10^5$ & $1.1\times10^{-8}$ \\
2189 &  5 &  4 & 0.8618 &  $-9413$ & $1.4\times10^5$ & $2.2\times10^{-8}$ \\
2194 & 11 &  9 & 0.8748 & $-10204$ & $2.8\times10^5$ & $1.1\times10^{-8}$ \\
2190 &  6 &  5 & 0.8856 & $-10899$ & $1.5\times10^5$ & $2.1\times10^{-8}$ \\
2191 &  7 &  6 & 0.9023 & $-12069$ & $1.5\times10^5$ & $2.0\times10^{-8}$ \\
2192 &  8 &  7 & 0.9148 & $-13023$ & $1.6\times10^5$ & $1.9\times10^{-8}$ \\
2193 &  9 &  8 & 0.9245 & $-13819$ & $1.7\times10^5$ & $1.8\times10^{-8}$ \\
2194 & 10 &  9 & 0.9322 & $-14494$ & $1.8\times10^5$ & $1.7\times10^{-8}$ \\
2195 & 11 & 10 & 0.9384 & $-15075$ & $1.9\times10^5$ & $1.6\times10^{-8}$ \\
2196 & 12 & 11 & 0.9436 & $-15580$ & $1.9\times10^5$ & $1.6\times10^{-8}$ \\
{\bf 2186} & {\bf 1} & {\bf  1} & {\bf 1.0000} & $\mathbf{-22947}$   & $\mathbf{9.2\times10^3}$ 
& $\mathbf{3.3\times10^{-7}}$ \\
2197 & 11 & 12 & 1.0597 & $-39074$ & $4.2\times10^4$ & $7.4\times10^{-8}$ \\
2196 & 10 & 11 & 1.0656 & $-41704$ & $3.4\times10^4$ & $9.1\times10^{-8}$ \\
2195 &  9 & 10 & 1.0728 & $-45365$ & $2.6\times10^4$ & $1.2\times10^{-7}$ \\
{\bf 2194} & {\bf 8} & {\bf  9} & {\bf 1.0817} & $\mathbf{-50811}$   & $\mathbf{1.9\times10^4}$ 
& $\mathbf{1.6\times10^{-7}}$ \\
2193 &  7 &  8 & 1.0931 & $-59775$ & $1.2\times10^4$ & $2.5\times10^{-7}$ \\
{\bf 2192} & {\bf 6} & {\bf  7} & {\bf 1.1082} & $\mathbf{-77314}$   & $\mathbf{6.6\times10^3}$ 
& $\mathbf{4.8\times10^{-7}}$ \\
{\bf 2191} & {\bf 5} & {\bf  6} & {\bf 1.1292} & $\mathbf{-127134}$  & $\mathbf{2.1\times10^3}$ 
& $\mathbf{1.6\times10^{-6}}$ \\
{\bf 2196} & {\bf 9} & {\bf 11} & {\bf 1.1431} & $\mathbf{-215975}$  & $\mathbf{1.4\times10^3}$ 
& $\mathbf{2.6\times10^{-6}}$ \\
{\bf 2190} & {\bf 4} & {\bf  5} & {\bf 1.1604} & $\mathbf{-1366152}$ & $\mathbf{1.6\times10^1}$ 
& $\mathbf{3.0\times10^{-4}}$ \\
\hline
\end{tabular}
\end{table}

In Table~\ref{resonances} the columns shows, from left to right: the
year of impact of the potential VI; the associated mean motion
resonance; the value of the resonant post-2185 semimajor axis $a'$;
the $\zeta$ coordinate of the keyhole center;
$\partial\zeta''/\partial\zeta$, i.e., the partial derivative of the
$\zeta$ coordinate on the post-2185 $b$-plane, taken with respect to
the $\zeta$ coordinate on the 2185 $b$-plane; and finally an estimate
of the maximum possible impact probability $P_{max}$ for the potential
VI in question.  Both $\zeta$ and $\partial\zeta''/\partial\zeta$ are
computed according to \citep{analytic}; in practice,
$\partial\zeta''/\partial\zeta$ can be seen as the factor by which the
stretching increases in the interval of time between the first and the
second encounter.

The values of $P_{max}$ are computed by multiplying the PDF by the
maximum possible chord (i.e., the diameter of the circle of radius
$b_\oplus$), and thus has to be seen as an upper limit; in this
respect, it should not be considered too surprising that the potential
VIs in the two top rows of Table~\ref{resonances} are not found by
either of the numerical procedures described in the paper, since it
may well be that the real values of the probability are significantly
smaller than $P_{max}$ because of small chords.  In the same spirit,
the good agreement between the values of $P_{max}$ in
Table~\ref{resonances} and those in the risk tables should not be
overestimated because of the very simple dynamical model with which the
analytical estimates are computed.
\end{appendix}

\end{document}